\documentclass[12pt]{article}
\usepackage{amsmath}
\usepackage{amsthm}
\usepackage{amssymb}
\usepackage{latexsym}
\usepackage{subeqnarray}
\usepackage{graphicx}
\usepackage{amsfonts}
\newcommand{\nc}{\newcommand*}

\theoremstyle{plain}
\newtheorem{thm}{Theorem}[section]
\nc{\reff}[1]{(\ref{#1})}
\nc{\nn}{\nonumber}
\nc{\ts}{\textstyle}
\nc{\ds}{\displaystyle}
\def\polP{\left\{P_n(x)\right\}_{n=0}^{\infty}}
\def\polQ{\left\{Q_n(x)\right\}_{n=0}^{\infty}}
\def\polpsi{\left\{\Psi_n(x)\right\}_{n=0}^{\infty}}
\usepackage[cp1251]{inputenc}
\inputencoding{cp1251}
\usepackage[T2B]{fontenc}
\usepackage[russian,english]{babel}
\begin{document}

{\bf V.V. Borzov \footnote{St.Petersburg State University of Telecommunications, 191065, Moika  61, St.Petersburg,
Russia. E-mail: borzov.vadim@yandex.ru},
E.V. Damaskinsky \footnote{ VI(IT) affilated to VA MTO, Russia, 191123, Zacharievskaya 22, St.Petersburg,
Russia. E-mail: evd@pdmi.ras.ru}}
\bigskip

\centerline{\large\bf  Invariance of the generalized oscillator}
\bigskip

\centerline{\large\bf under linear transformation of}
\bigskip

\centerline{\large\bf the related system of orthogonal polynomials}
\bigskip

\begin{quote}
We consider two families of polynomials $\mathbb{P}=\polP$ and $\mathbb{Q}=\polQ$\footnote{Here and below
we consider only monic polynomials.} orthogonal on the real line
  with respect to probability measures $\mu$ and $\nu$ respectively. Let $\polQ$ and $\polP$ connected by the linear relations
$$ Q_n(x)=P_n(x)+a_1P_{n-1}(x)+...+a_kP_{n-k}(x).$$
Let us denote $\mathfrak{A}_P$ and $\mathfrak{A}_Q$
generalized oscillator algebras  associated with the sequences $\mathbb{P}$ and $\mathbb{Q}$.
In the case $k=2$  we describe all pairs ($\mathbb{P}$,$\mathbb{Q}$), for which
the algebras $\mathfrak{A}_P$ and $\mathfrak{A}_Q$ are equal.
In addition, we construct corresponding algebras of generalized oscillators for arbitrary $k\geq1$.
\end{quote}

\section{Introduction}
Let $\mathbb{P}=\polP$ is a family of polynomials orthogonal on the real line
 with respect to the probability measure $\mu$. Consider the sequence of polynomials
$\mathbb{Q}=\polQ$ such that
\begin{equation*}
    Q_n(x)=P_n(x)+a_1P_{n-1}(x)+...+a_kP_{n-k}(x), \quad n>k-1.
\end{equation*}
The family of orthogonal polynomials associated with such linear relation was discussed in several works
(see e.g. \cite{1} - \cite{5}).
In particular the necessary and sufficient conditions for the orthogonality of the sequence
$\polQ$ with respect to a positive Borel measure $\nu$ on the real line are given in the article \cite{5}.

It is known \cite{6} that every sequence of polynomials $\polP$
 orthogonal with respect to positive Borel measures $\mu$ on the real line generates
the generalized oscillator algebra $\mathfrak{A}_P$. In this work, we investigate the question under what
conditions algebras $\mathfrak{A}_P$ and $\mathfrak{A}_Q$,
generated by such linearly related polynomials, coincide
\begin{equation*}
\mathfrak{A}_P=\mathfrak{A}_Q.
\end{equation*}
 This problem was considered in \cite{8} for the simplest case $k=1$. In this paper we discuss the case $k=2$.

Below we will need the following results (\cite{5}-\cite{7}).
Let $u$ is a linear functional on the linear space of polynomials with real coefficients.
The polynomials $\polP$ are called orthogonal with respect to $u$, if
$$
\langle u, P_nP_m\rangle=0,\, \forall n\neq m \quad\text{and}\quad
\langle u, P_n^{\,\,2}\rangle\neq 0,\, \forall n=0,1,\ldots .
$$
Let $H=\left\{ u_{i+j}\right\}_{i,\,j\geq 0}$
is the Hankel matrix associated with the functional
 $u$, where $u_k=\langle u, x^{k}\rangle,k\geq 0$.
The linear functional $u$ is called the quasi-definite (positive definite)
functional, if the leading
submatrices $H_n$ of the matrix $H$ are nonsingular (positive definite) for all $n\geq0$.

Favard's theorem gives a description of quasi-definite (positively definite) linear functional
in terms of the three-term  recurrence relations that satisfied by the polynomials  $\polP$:
\begin{equation*}
xP_{n}(x)=P_{n+1}(x)+\beta_n P_{n}(x)+\gamma_nP_{n-1}(x)
\end{equation*}
\begin{equation*}
   P_{0}(x)=1,\qquad P_{1}(x)=x-\beta_0,
\end{equation*}
where $\gamma_n\neq0$ (respectively $\gamma_n>0$).

If $u$ is a positive definite linear functional, then there exists the positive Borel measure $\mu$
(supported by an infinite subset of $\mathbb{R}$) such that
\begin{equation*}
\langle u, q\rangle=\int_{\mathbb{R}} q \rm{d}\mu, \qquad \forall q\in\mathbb{R}.
\end{equation*}

Bellow we will use the following theorem.
\begin{thm}\label{38}\,{\rm \cite{5}}
Let $\polP$ is the sequence of orthogonal polynomials with recurrence relations
\begin{equation}\label{01}
xP_n(x)=P_{n+1}(x)+\beta_nP_{n}(x)+\gamma_nP_{n-1},\quad (\gamma_n\neq0)
\end{equation}
\centerline{$P_0(x)=1,\quad P_1(x)=x-\beta_0.$}
Let $a_1$ and $a_2$ are real numbers such that $a_2\neq0$, and $Q_n(x)$ are polynomials
defined by the relations
\begin{equation}\label{02}
Q_n(x)=P_n(x)+a_1P_{n-1}(x)+a_2P_{n-2},\quad n\geq 3.
\end{equation}
Then the orthogonality of the sequence $\polQ$ depends on the choice of the coefficients $a_1$ and $a_2$.
More precisely, $\polQ$ is a family of orthogonal polynomials with recurrence relations
\begin{equation}\label{03}
xQ_n(x)=Q_{n+1}(x)+\widetilde{\beta}_nQ_{n}(x)+\widetilde{\gamma}_nQ_{n-1},\quad n\geq1,\quad
(\gamma_n\neq0)
\end{equation}
if and only if \qquad $\gamma_3+a_1(\beta_2-\beta_3)\neq0$\quad and
\medskip

(${i}$) if $a_1=0$, then for $n\geq4$,\quad $\beta_n=\beta_{n-2}$,\quad  $\gamma_n=\gamma_{n-2}$;
\medskip

(${ii}$) if $a_1\neq0$ and $a_1^{\,\,2}=4a_2$, then for $n\geq2$
$$
\beta_n=A+Bn+Cn^2,\quad \gamma_n=D+En+Fn^2,
$$
with \quad$a_1C=2F$, $a_1B=2E-2F$,\qquad $(A,B,C,D,E,F\in\mathbb{R})$;
\medskip

(${iii}$)  if $a_1\neq0$ and $a_1^{\,\,2}>4a_2$, then for $n\geq2$
$$
\beta_n=A+B\lambda^n+C\lambda^{-n},\quad \gamma_n=D+E\lambda^n+F\lambda^{-n}
$$
with \quad $a_1C=(1+\lambda)F$, $a_1\lambda B=(1+\lambda)E$,\quad $(A,B,C,D,E,F\in\mathbb{R})$,
where $\lambda$ is the unique solution in $(-1,1)$ of the equation

\qquad\qquad\qquad $a_1^{\,\,2}\lambda=a_2(1+\lambda)^2$;

(${iv}$) if $a_1\neq0$ and $a_1^{\,\,2}<4a_2$, and we let $\lambda=e^{i\theta}$, $\theta\in (0,\pi)$
be the unique solution of the equation $a_1^{\,\,2}\lambda=a_2(1+\lambda)^2$, then for $n\geq2$
$$
\beta_n=A+Be^{in\theta}+\overline{B}e^{-in\theta},\quad
\gamma_n=D+Ee^{in\theta}+\overline{E}e^{-in\theta},
$$
with $a_1\lambda B=(1+\lambda)E$\quad $(A,D\in\mathbb{R},\, B,F\in\mathbb{C})$.
\end{thm}
\bigskip

Let us give the definition of the generalized oscillator connected with the
family of orthogonal polynomials \cite{6}. Let $\mu$ is a probability measure on $\mathbb{R}$ with
finite moments of all orders
\begin{equation*}
    \mu_n=\int_{-\infty}^{+\infty} x^n\rm{d}\mu <\infty, \quad n\geq0.
\end{equation*}
These moments  define uniquely two sequences of real numbers
$\left\{a_n\right\}_{n=0}^{\infty}$,
$\left\{b_n\right\}_{n=0}^{\infty}$ and  the system of orthogonal polynomials $\polpsi$ which satisfy recurrence relations
\begin{equation}\label{06}
x\Psi_{n}(x)=b_n\Psi_{n+1}(x)+a_n \Psi_{n}(x)+b_{n-1}\Psi_{n-1}(x),
\end{equation}
for $n\geq0$ and also initial conditions
\begin{equation*}
   \Psi_{0}(x)=1,\qquad \Psi_{1}(x)=\frac{x-a_0}{b_0}.
\end{equation*}
These polynomials form an orthonormal basis in the Hilbert space
$\mathcal{H}=\rm{L}_2(\mathbb{\mathbb{R}};\mu)$.

It is necessary to distinguish two cases
\begin{enumerate}
  \item $a_n=0$ - symmetric case;
  \item $a_n\neq0$ - nonsymmetric case.
\end{enumerate}

In the Hilbert space $\mathcal{H}$ we define the ladder operators $a^{\pm}$ and the number operator $N$
by the formulas
\begin{align*}
a^+\Psi_{n}(x)&=\sqrt{2}b_n\Psi_{n+1}(x),\\
a^-\Psi_{n}(x)&=\sqrt{2}b_{n-1}\Psi_{n-1}(x),\\
N\Psi_{n}(x)&=n\Psi_{n}(x),\quad n\geq0.
\end{align*}
Let $B(N)$ be an operator-valued function such that
\begin{align*}
B(N)\Psi_{n}(x)&={b_{n-1}}^2\Psi_{n}(x), \\
B(N+\mathbb{I})\Psi_{n}(x)&={b_{n}}^2\Psi_{n}(x),\quad n\geq0.
\end{align*}

The next theorem is faithful.
\begin{thm}\label{07}\, {\rm \cite{6}}\,
The operators $a^{\pm}$, $N$, $\mathbb{I}$ satisfy the following relations
\begin{equation}
a^-a^+=2B(N+\mathbb{I}), \quad a^+a^-=2B(N), \quad
[N,a^{\pm}]=\pm a^{\pm}.
\end{equation}
\end{thm}

\noindent {\bf Definition}.\quad {\it The associative algebra
$\mathfrak{A}_{\Psi}$ generated by the operators
$a^{\pm}$, $N$, $\mathbb{I}$ satisfying the relations of {\bf theorem 1.2} and by
the commutators of these operators is called the generalized oscillator algebra corresponding
to the orthonormal system of
polynomials $\polpsi$ with recurrence relations}  \reff{06}.
\bigskip

We will give one useful consequence of the previous theorem.
Let $\mathfrak{A}_s$ is the algebra of generalized oscillator corresponding to
recurrence relations \reff{06} in the symmetric case ($a_n=0$) and $\mathfrak{A}_a$ is the algebra
of generalized oscillator corresponding recurrence relations \reff{06} in an asymmetric case
($a_n\neq0$). Then $\mathfrak{A}_s=\mathfrak{A}_a$.

Now we are ready to formulate the problem under consideration.
We suppose that there are two families of polynomials $\mathbb{P}=\polP$ and $\mathbb{Q}=\polQ$
orthogonal with respect to probability measures $\mu$ and $\nu$ respectively.
We suppose that these polynomials satisfy the conditions of {\bf theorem 1.1}
and
$\beta_n, \widetilde{\beta_n}, \gamma_n, \widetilde{\gamma_n}\in \mathbb{R}$.
Let us denote $\mathfrak{A}_P$ and $\mathfrak{A}_Q$ the corresponding algebra of generalized oscillators.
\bigskip

\noindent{\bf Problem}. {\it We want to describe all pairs of families of orthogonal polynomials
$(\mathbb{P}, \mathbb{Q})$ for which $\mathfrak{A}_P=\mathfrak{A}_Q$}.

\section{Jacobi matrices and the main result}
Let $\mathbb{P}$ and $\mathbb{Q}$ are defined in Hilbert spaces
$\mathcal{H}_{\mu}=\rm{L}^2(\mathbb{R};\mu)$ and $\mathcal{H}_{\nu}=\rm{L}^2(\mathbb{R};\nu)$,
respectively. Let these polynomials satisfy the recurrence relations \reff{01} and \reff{03},
respectively. In addition, we assume that these polynomials are related
to each other
by the relation \reff{02}.
We will also suppose that
$\beta_n, \widetilde{\beta_n}, \gamma_n, \widetilde{\gamma_n}\in \mathbb{R}$ and
the coefficients $\beta_n, \gamma_n$ satisfies the conditions of {\bf theorem 1.1}.

The Jacobi matrices $J_P$, $J_Q$ corresponding to the RR \reff{01}, \reff{03},
respectively, can be written in the form
$$
J=\begin{bmatrix} A&I_1\\ I_2&B\end{bmatrix},\quad
I_1=\begin{bmatrix} 0&0&0&\cdots\\ 0&0&0&\cdots\\ 1&0&0&\cdots\end{bmatrix},\quad
I_2=\begin{bmatrix}
0&0&\gamma_3\\ 0&0&0\\ 0&0&0\\ \cdots&\cdots&\cdots
\end{bmatrix}
$$
where the matrix $A$ for sequences $\mathbb{P}$  and $\mathbb{Q}$ have the following form
$$
A_P=\begin{bmatrix}
\beta_0&1&0\\
\gamma_1&\beta_1&1\\
0&\gamma_2&\beta_2
\end{bmatrix},\quad
A_Q=\begin{bmatrix}
\widetilde{\beta}_0&1&0\\
\widehat{\gamma}_1&\widetilde{\beta}_1&1\\
0&\widetilde{\gamma}_2&\widetilde{\beta}_2
\end{bmatrix},
$$
and the matrix $B$ equals to
$$
B=\begin{bmatrix}
\beta_3&1&0&0&0&0&\cdots\\
\gamma_2&\beta_2&1&0&0&0&\cdots\\
0&\gamma_3&\beta_3&1&0&0&\cdots\\
0&0&\gamma_2&\beta_2&1&0&\cdots\\
\cdots&\cdots&\cdots&\cdots&\cdots&\cdots&\cdots
\end{bmatrix}.
$$

Let us note that elements $(\beta_0,\beta_1,\beta_2,\beta_3)$ and $(\gamma_1,\gamma_2,\gamma_3)$ of the
matrix $J_P$ are given, while the elements $(\widetilde{\beta}_0,\widetilde{\beta}_1,\widetilde{\beta}_2)$
as well as coefficients $a_1,a_2$ in the equation \reff{02} should be defined.
\bigskip

We will consider all 4 cases of the {\bf theorem 1.1}. General relations valid for all four cases have the following form:

\begin{align}\label{11}
\widetilde{\gamma}_n&=\gamma_n, \, n\geq1;&\quad & \widetilde{\beta}_n=\beta_n, \, n\geq3; \nonumber \\
\beta_{2n+1}&=\beta_3, \, n\geq1;&\quad & \gamma_{2n}=\gamma_2, \, n\geq1; \\
\beta_{2n}&=\beta_2, \, n\geq1;&\quad & \gamma_{2n+1}=\gamma_3, \, n\geq1; \nonumber \\
&\gamma_{n}\neq 0;&\quad & \qquad a_2\neq 0. \nonumber
\end{align}

We use the following notation
\begin{gather}\label{15}
s_1=\frac{\gamma_3-\gamma_1-(\beta_2-\beta_1)(\beta_3-\beta_1)}{\gamma_3},
\nonumber \\
s_2=\frac{\gamma_2}{\beta_3-\beta_1},\quad s_3=\frac{\gamma_3}{\beta_3-\beta_1},\\
w=\frac{a_1}{4s_3}-\frac{\gamma_2}{\gamma_3},\quad
w_{\lambda}=\frac{\lambda}{(1+\lambda)^2}\, \frac{a_1}{s_3}-\frac{\gamma_2}{\gamma_3}.\nonumber
\end{gather}

We now formulate our main result in terms of Jacobi matrices.
Namely, we can prove that all pairs of orthogonal polynomial  systems $\polP$ and $\polQ$ connected by
the linear relation \reff{02}, which generate the same algebra of generalized oscillator,
can be divided into following eight groups:
\medskip

{\bf The case I}\quad $a_1=0$,\, $\beta_1\neq\beta_3$;

In this case the matrix $A_Q$ and the coefficient $a_2$ are defined uniquely by the relations
\begin{gather}\label{16}
\widetilde{\beta}_0=\beta_0+\frac{(\beta_3-\beta_1-\beta_0)\gamma_1}{\gamma_2\gamma_3}\,a_2,\quad
a_2=-s_1s_3^{\,\,2},\nn \\
\widetilde{\beta}_1=\beta_1+\frac{a_2}{s_3}-\widetilde{\beta}_0,\quad \widetilde{\beta}_2=- \frac{a_2}{s_3},\\
\widetilde{\gamma}_n=\gamma_n,\, n\geq1, \quad \widetilde{\beta}_n=\beta_n,\, n\geq3.\nn
\end{gather}
\medskip

{\bf The case II}\quad $a_1=0$,\, $\beta_1=\beta_3$,\, $\gamma_3=\gamma_1$,\, $\beta_2\neq\beta_0$.

In this case the matrix $A_Q$ and the coefficient $a_2$ are defined uniquely by the relations
\begin{gather}\label{17}
\widetilde{\beta}_0=\beta_1-\frac{\gamma_2}{\beta_2-\beta_0},\quad
\widetilde{\beta}_1=\beta_0+\frac{\gamma_2}{\beta_2-\beta_0},\nn \\
\widetilde{\beta}_2=\beta_2,\, a_2=\gamma_2\,\frac{\beta_1-\beta_0}{\beta_2-\beta_0}-
\frac{\gamma_2^{\,\,2}}{(\beta_2-\beta_0)^2},\\
\widetilde{\gamma}_n=\gamma_n,\, n\geq1, \quad \widetilde{\beta}_n=\beta_n,\, n\geq3. \nn
\end{gather}
\medskip

{\bf The case III}\quad  $a_1\neq0$,\, $\beta_1\neq\beta_3$,\, $a_2=\frac{1}{4}a_1^{\,\,2}$.

We denote by $w$ the solution of the equation
\begin{multline*}
\qquad 16s_3^{\,\,2}w^4+32s_2s_3w^3+(16s_2^{\,\,2}+4s_3^{\,\,2})w^2+\\
(4s_2+s_1s_3)w+s_1s_2+\frac{\gamma_2}{\gamma_3}(\beta_2-\beta_1)=0 \qquad
\end{multline*}
such that $a_1=(4s_3w+4s_2)\in\mathbb{R}$ and introduce the quantity
\begin{multline}\label{08}
C_{\beta,\gamma}=a_2\left[- \frac{\gamma_1}{\gamma_3}\left(\beta_2-a_1(w+1)\right)+\beta_0(4w^3+4w+1)\right]-\\
a_1w\left[\beta_0(\beta_1+\beta_2)+\gamma_1\right].\qquad\quad
\end{multline}
In this case, for given $w$, the matrix $A_Q$ and the coefficients $a_1,\,a_2$ are defined uniquely by the relations
\begin{gather}\label{09}
\widetilde{\beta}_0=\beta_0-\frac{C_{\beta,\gamma}}{\gamma_2},\quad
\widetilde{\beta}_1=\beta_1+\frac{C_{\beta,\gamma}}{\gamma_2}+a_1w,\nn\\
\widetilde{\beta}_2=\beta_2-a_1(w+1),\\
a_1=4s_3w+4s_2,\quad a_2=\frac{1}{4}a_1^{\,\,2}.
\end{gather}
\medskip

{\bf The case IV}\quad  $a_1\neq0$,\, $\beta_1=\beta_3$,\, $a_2=\frac{1}{4}\,a_1^{\,\,2}$.

Let $a_1$ be a real solution of the equation
\begin{equation*}
\frac{\gamma_2^{\,\,2}}{\gamma_3^{\,\,2}}\,a_1^{\,\,2}-
a_1\left(\frac{\gamma_2}{\gamma_3}+\frac{\gamma_1}{4\gamma_3}-\frac{1}{4}\right)+
\frac{\gamma_2}{\gamma_3}(\beta_2-\beta_1)=0.
\end{equation*}
In this case, the matrix $A_Q$ is defined uniquely by the relations
\begin{gather}\label{10}
\widetilde{\beta}_0=\beta_0-\frac{D_{\beta,\gamma}}{\gamma_2},\quad
\widetilde{\beta}_1=\beta_1+\frac{D_{\beta,\gamma}}{\gamma_2}+a_1w,\\
\widetilde{\beta}_2=\beta_2-a_1(w+1), \nn
\end{gather}
where $w=- \ds\frac{\gamma_2}{\gamma_3}$ and $D_{\beta,\gamma}$ is defined by \reff{08} at
$w=- \ds\frac{\gamma_2}{\gamma_3}$.
\medskip

{\bf The case V}\quad   $a_1\neq0$,\,  $a_1^{\,\,2}>4a_2$,\, $a_2=\ds\frac{\lambda}{(1+\lambda)^2}\,a_1^{\,\,2}$,\,
$\beta_1\neq\beta_3$,\, $\lambda\in(-1,1)$.

Let $w_{\lambda}$ be a solution of the equation
\begin{multline}\label{18}
\frac{(1+\lambda)^4}{\lambda^2}\,s_3^{\,\,2}w_{\lambda}^{\,\,4}+
2\frac{(1+\lambda)^2}{\lambda}\,s_2s_3w_{\lambda}^{\,\,3}+\\
\left(\frac{(1+\lambda)^4}{\lambda^2}\,s_2^{\,\,2}+\frac{(1+\lambda)^2}{\lambda}\,s_3\right)w_{\lambda}^{\,\,2}+\\
\left(\frac{(1+\lambda)^2}{\lambda}\,s_2+s_1s_3\right)w_{\lambda}+s_1s_2+\frac{\gamma_2}{\gamma_3}(\beta_2-\beta_1)=0,
\end{multline}
such that
$$
a_1=\frac{(1+\lambda)^2}{\lambda}(s_3w_{\lambda}+s_2)\in\mathbb{R}.
$$
In this case, the matrix $A_Q$ is defined uniquely by the relations
\begin{gather}\label{12}
\widetilde{\beta}_0=\beta_0-\frac{C_{\lambda}}{\gamma_2},\quad
\widetilde{\beta}_1=\beta_1+\frac{C_{\lambda}}{\gamma_2}+a_1w_{\lambda},\quad
\widetilde{\beta}_2=\beta_2-a_1(w_{\lambda}+1),
\end{gather}
where
\begin{multline}\label{13}
C_{\lambda}\!=\!a_2\left[-\frac{\gamma_1}{\gamma_3}\left(\beta_2-a_1(w_{\lambda}\!+\!1)\right)+
\beta_0\left(\frac{(1+\lambda)^2}{\lambda}(w_{\lambda}^{\,\,2}+w_{\lambda})+1\right)\right]\\
-a_1w_{\lambda}\left[\beta_0(\beta_1+\beta_2)+\gamma_1\right]\qquad\quad
\end{multline}
$$
a_1=\frac{(1+\lambda)^2}{\lambda}(s_3w_{\lambda}+s_2),\quad a_2=\frac{\lambda}{(1+\lambda)^2}a_1^{\,\,2},
$$
and $\lambda$ is a free parameter.
\medskip

{\bf The case VI}\quad   $a_1\neq0$,\,  $a_1^{\,\,2}>4a_2$,\, $a_2=\ds\frac{\lambda}{(1+\lambda)^2}\,a_1^{\,\,2}$,\,
$\beta_1=\beta_3$,\, $\lambda\in(-1,1)$.

Let $a_1$ be a real solution of the equation
\begin{equation*}
\frac{\gamma_2^{\,\,2}}{\gamma_3^{\,\,2}}a_1^{\,\,2}-
a_1\left[\frac{\gamma_2}{\gamma_3}+\frac{\lambda}{(1+\lambda)^2}\left(\frac{\gamma_1}{\gamma_3}-1\right)\right]+
\frac{\gamma_2}{\gamma_3}(\beta_2-\beta_1)=0,
\end{equation*}
$$
a_2=\frac{\lambda}{(1+\lambda)^2}a_1^{\,\,2}.
$$
In this case, the matrix $A_Q$ is defined uniquely by the relations
\begin{gather}\label{14}
\widetilde{\beta}_0=\beta_0-\frac{D_{\lambda}}{\gamma_2},\quad
\widetilde{\beta}_1=\beta_1+\frac{D_{\lambda}}{\gamma_2}+a_1w_{\lambda},\quad
\widetilde{\beta}_2=\beta_2-a_1(w_{\lambda}+1),
\end{gather}
where $w_{\lambda}=-\ds\frac{\gamma_2}{\gamma_3}$ and $D_{\lambda}$ is defined by the relation \reff{13} for
$C_{\lambda}$ with $w_{\lambda}=-\ds\frac{\gamma_2}{\gamma_3}.$
\medskip

{\bf The cases VII and VIII}\quad $a_1\neq0$,\,  $a_1^{\,\,2}<4a_2$,\,
$a_2=\ds\frac{\lambda}{(1+\lambda)^2}\,a_1^{\,\,2}$,\, \, $\lambda=e^{i\theta}$,\, $\theta\in (0,\pi).$

The other relations in the case VII are the same as in the case V, and  in the case VIII are the same as
in the case VI.

\section{ Possibility of generalization to the case of $k>2$
}
We give  few comments about  construction all possible pairs of polynomial systems connected
by the general linear relation
\begin{equation}\label{19}
    Q_n(x)=P_n(x)+a_1P_{n-1}(x)+\ldots+a_kP_{n-k},\quad  k\geq 2
\end{equation}
for which $\mathfrak{A}_P=\mathfrak{A}_Q$.

From results of the paper \cite{5} we have

1)$\widetilde{\gamma}_n=\gamma_n$,\quad $n\geq1$;

2)$a_1,a_2,\ldots,a_k\in\mathbb{R}$,\, $a_k\neq0$,\, $\gamma_n\neq0$,\, $n\geq1$;

3) $\widetilde{\beta}_n=\beta_n$,\, $\widetilde{\gamma}_n=\gamma_n+a_1(\beta_{n-1}-\beta_n)\neq0$,\quad $n\geq k+1$;

4) $\gamma_n-\gamma_{n-k}=a_1(\beta_{n-1}-\beta_n)$, \, $n\geq k+2$.
\medskip

You should  consider two alternative cases:
$$
A)\quad a_1=0;\qquad B)\quad a_1\neq0,\, \beta_n=\beta_k,\, n\geq k.
$$
In both cases we have $\gamma_n=\gamma_{n-k},\, n\geq k+2$.

Then we have to consider
all possible cases when some of the coefficients
$a_1,a_2,\ldots,a_{k-1}$ are not equal to zero and all others equal to zero.
We recall that $a_k\neq0$. The number of such variants is equal to
$$
C_{k-1}^0+C_{k-1}^1+\cdots+C_{k-1}^{k-1}=2^{k-1}.
$$

\section{The generalized oscillator algebra corresponding to the pair $(\mathbb{P},\mathbb{Q})$}

We discuss                                                                                                                                                                                                  the generalized oscillator  algebra
$\mathfrak{A}_\mathbb{P}=\mathfrak{A}_\mathbb{Q}$
for considered above pair $(\mathbb{P},\mathbb{Q})$ of orthogonal polynomials  systems
connected with each other by the linear relation \reff{19}.
For the construction of the
 generalized oscillator algebra $\mathfrak{A}_\mathbb{P}$ in the Hilbert space
$\mathcal{H}_\mu=\text{L}^2(\mathbb{R};\mu)$  should be symmetrize the recurrence relations \reff{01}
(see \cite{6})
\begin{equation}\label{41}
xP_{n}(x)=P_{n+1}(x)+\beta_nP_{n}(x)+\gamma_nP_{n-1}(x),\, P_0=1,
\end{equation}
with $\gamma_0=0$.

With this purpose let us proceed from the system of polynomials $\mathbb{P}$ to the system of polynomials
$\Phi=\left\{\varphi_n(x)\right\}_{n=0}^{\infty}$ by the formulas
\begin{equation*}
P_n(x)=\alpha_n\varphi_n(x),\,\, \alpha_0=1,\, \alpha_n=\sqrt{\gamma_1}(\gamma_2...\gamma_s)^{\,\,\frac{m}{2}} (\gamma_{s+1}...\gamma_{k+1})^{\,\,\frac{m-1}{2}},
\end{equation*}
where $n=(m-1)k+s,\, m\geq 1,\, s=2,...k+1$ and $k\geq 1$ is a fixed integer.
This is only possible if $\gamma_1\neq0,...,\gamma_{k+1}\neq0 $. Then recurrence relations
\reff{41} take the form
\begin{equation*}
x\varphi_{n}(x)=b_{n}\varphi_{n+1}(x)+\beta_n\varphi_{n}(x)+b_{n-1}\varphi_{n-1}(x),
\end{equation*}
where
\begin{equation*}
\varphi_{0}=1,\quad b_{-1}=0,\quad b_{n}=\sqrt{\gamma_{n+1}},\quad n\geq0,
\end{equation*}
and
\begin{gather}\label{44}
\gamma_n= \left\{\begin{aligned}\gamma_1&\quad\text{if}\quad n=1,\\
\gamma_s&\quad\text{if}\quad n=km+s,\quad m\geq0,\quad s=2,...{k+1}.\end{aligned}\right.
\end{gather}

We define the ladder operators $a^{\pm}_{\Phi}$ and  the number operator $\mathbb{N}_\Phi$ in
$\mathcal{H}_\mu$ by formulas
\begin{align*}
a^{+}_{\Phi}\varphi_n=\sqrt{2\gamma_{n+1}}\varphi_{n+1},\\
a^{-}_{\Phi}\varphi_n=\sqrt{2\gamma_{n}}\varphi_{n-1},\\
\mathbb{N}_\Phi \varphi_n=n\varphi_n,\quad n\geq0.
\end{align*}
Let $B_\Phi(\mathbb{N}_\Phi)$ be an operator-valued function defined by the following
equalities
\begin{equation*}
\quad B_\Phi(\mathbb{N}_\Phi)\varphi_n=\gamma_n\varphi_n,\quad n\geq0.
\end{equation*}
\begin{equation*}
B_\Phi(\mathbb{N}_\Phi+I_\mu)\varphi_n=\gamma_{n+1}\varphi_n,\quad n\geq0.
\end{equation*}

Then the generalized oscillator  algebra $\mathfrak{A}_{\mathbb{P}}$ is generated by the operators $a^{\pm}_{\Phi}$,
$\mathbb{N}_\Phi$, $I_\mu$ satisfying the relations
\begin{equation}\label{21}
\left\{
\begin{aligned}
&a^{-}_{\Phi}a^{+}_{\Phi}=2B_\Phi(\mathbb{N}_\Phi+I_\mu),\quad
a^{+}_{\Phi}a^{-}_{\Phi}=2B_\Phi(\mathbb{N}_\Phi),\\
&\qquad\qquad\qquad [\mathbb{N}_\Phi, a^{\pm}_{\Phi}]=\pm a^{\pm}_{\Phi}
\end{aligned}
\right.
\end{equation}
and by the commutators of these operators.

In this case, the quadratic Hamiltonian
\begin{equation*}
H_\Phi = a^{-}_{\Phi}a^{+}_{\Phi}+a^{+}_{\Phi}a^{-}_{\Phi}
\end{equation*}
is a selfadjoint operator in the Hilbert space $\mathcal{H}_\mu$.

Orthonormal polynomials $\left\{\varphi_n(x)\right\}_{n=0}^{\infty}$
are eigenfunctions of the operator $H_\Phi$. The corresponding eigenvalues are equal to
\begin{equation*}
\lambda_0=2\gamma_1,\quad
\lambda_n=2(\gamma_n + \gamma_{n+1}),\quad n\geq1.
\end{equation*}

In conclusion, let us note that since in our case
$$
b_n^{\,\,2}\neq (a_0+a_2n)(1+n),
$$
then according to the results of \cite{9}, \cite{10}
$$
\dim\mathfrak{A}_\mathbb{P}=\dim\mathfrak{A}_\mathbb{Q}=\infty .
$$

Note also that it would be interesting to study the relation of the orthogonality measures  $\mu$ and $\nu$
in the spaces $\mathcal{H}_\mu$ and $\mathcal{H}_\nu$ respectively.

\smallskip

{\bf Acknowledgment.}\, EVD grateful to RFBR for financial support under the grant 15-01-03148.


\smallskip
\begin{thebibliography}{19}

\bibitem{1} V.B.Uvarov, {\it The connection between systems of polynomials orthogonal with respect to
different distribution function},  U.S.S.R. Comput. Math.and Math. Phys., {\bf 9}, no.6, 25-36 (1969).
\bibitem{2} K.H. Kwon, J.H. Lee, F. Marcellan, {\it Orthogonality of linear combinations of two orthogonal sequences},
Journ. Comput. Appl. Math., {\bf 137}, 109-122 (2001).
\bibitem{3} E. Berriochoa, A. Cachafiero, J.M. Garcia-Amor, {\it A characterization of the four Chebyshev
orthogonal families}, Int. J. Math and Math. Sci. 2005:13, 2071-2079 (2005).
\bibitem{4} Z.S. Grinshpun, {\it Differential equation for the Bernstein-Szego orthogonal polynomials},
Differ.Equ. {\bf 26}, no.5, 545-550 (1990).
\bibitem{5} M. Alfaro, F. Marcellan, Ana Pena, M.L. Rezola, {\it When do linear combinations of orthogonal polynomials
yield new sequences of orthogonal polynomials?}, J. Comput. Appl. Math, {\bf 233}, 1146-1452 (2010).
\bibitem{6} V.V. Borzov, {\it Orthogonal polynomials and generalized oscillator algebras}, Integral Transf. and
Special Functions, {\bf 12}(2), 115-138 (2001).
\bibitem{7} T.S. Chihara, {\it An Introduction to Orthogonal Polynomials},
Gordon and Breach, New York, 1978.
\bibitem{8} V.V. Borzov, E.V. Damaskinsky, {\it On representations of generalized  oscillator for two sequences
 of linearly related orthogonal polynomials} Proc. Int. Conf. DAYS on DIFFRACTION 2015, pp.30-33 (2015).
\bibitem{9} G. Honnouvo, K. Thirulogasanthar, {\it On the dimensions of the oscillator algebras induced
by orthogonal polynomials} J. Math. Phys. {\bf 55}, 093511 (2014).
\bibitem{10} V.V. Borzov, E.V. Damaskinsky, {\it On dimensions of oscillator algebras} Proc. Int. Conf.
DAYS on DIFFRACTION 2014, pp.48-52 (2014).
\end{thebibliography}
\end{document}